# Monitoring and Regulation of Micro-Displacement Deviation in Few-Mode Beam Alignment through Mode Decomposition

Lin Xu, Li Pei, Jianshuai Wang, Zhouyi Hu, and Tigang Ning

*Abstract*—Beam alignment enables efficient, stable transmission and control of optical energy and information, which critically depend on precise monitoring and regulation of the three-dimensional (3D) relative positioning between fibers. This study introduces an approach to achieve more accurate 3D measurement of the spatial displacement between two optical fibers in a few-mode configuration, by integrating mode decomposition with a straightforward machine learning algorithm. This method leverages inherent information from the optical field, enabling precise beam alignment with a simple structure and minimal computational effort. In the 3D measurement experiment, the proposed method achieves a coefficient of determination of 0.99 for transverse offsets in the x- and y-directions, and 0.98 for air gap in the z-direction. The RMSE in x-direction, y-direction and z-direction is respectively 0.135 μm, 0.128 μm and 2.42 μm. The time for a single 3D displacement calculation is $4.037\times10^{-4}$ seconds. Furthermore, it facilitates single-step displacement regulation with a deviation tolerance within 0.15 μm and modal content regulation with an accuracy of 4.67%. These results establish a theoretical framework for addressing key challenges in optical path alignment, crosstalk compensation, precision instrument manufacturing, and fiber optic sensing.

## I. INTRODUCTION

WITH the rollout of 6G communication and services[1], sensors and controllers with integrated communication and sensing capabilities have attracted significant attention[2], [3], [4], [5]. Optical fiber, as the primary transmission medium in communication systems, is perfectly suited for building integrated communication and sensing systems using optical fibers as the main sensor components[6], [7]. Meanwhile, three-dimensional (3D) displacement sensing is a fundamental and crucial sensing technology[8], [9], [10], [11], [12], [13], [14], [15], [16], [17], [18]. It plays a vital role in precision manufacturing[12], automated monitoring[13], and free-space optical communication[14]. Research into 3D displacement sensing technology that utilizes communicable optical fibers as the primary sensor components is of great significance for the further development of the future communication industry and for monitoring and manufacturing in challenging environments. Currently common 3D displacement sensing technologies include time-of-flight[16], structured light[18], and dual-camera methods[11]. However, all have drawbacks such as high cost, complex equipment requirements, and high computational complexity. Compared with recent literature on 3D displacement measurement, Reference [17] from year of 2022 mentions achieving a simulation error of approximately 4 μm and an experimental error of approximately 8 μm in three directions using a capacitive grating sensor. Reference [15] from 2022 demonstrates a displacement measurement error of less than 3 μm in the x-direction and less than 6 μm in the z-direction using a single-camera method with moiré fringes. Reference [11] from 2025 optimizes the displacement measurement method based on generalized dual-camera relative motion estimation, achieving an average error of less than 0.46 mm in the x-direction and a root mean square error of less than 0.89 mm in all three directions. Reference [18] from 2023 explicitly mentions a measurement performed in 13.27 seconds with a transverse resolution of 2.7 μm and a longitudinal resolution of 37 μm. Although these literatures have different measurement ranges and algorithms lead to varying accuracies, the micro-displacement measurement accuracy in Reference [17] is still unsatisfactory. Therefore, a new mechanism is urgently needed to improve measurement accuracy and speed while reducing instrument complexity. At the same time, a combination of more detailed measurements and measurements of larger particles is needed to enable 3D displacement measurement to be applied to more scenarios. We also compared the accuracy of 2D measurements. The research in [19] is a 2D displacement measurement method and can control the error to approximately 0.07 μm, but this method still requires a telecentric camera. In addition, these offset distance adjustment methods all use cameras to measure alignment radially from the outside of the fiber. This requires the camera to simultaneously capture the cladding sidewalls of two

This work was supported in part by the National Key Research and Development Program of China under Grant 2024YFF0726401 and in part by the National Natural Science Foundation of China under Grant 62305020 and Grant 62401038. *(Corresponding authors: Li Pei.)*

Lin Xu, Li Pei, Jianshuai Wang, Zhouyi Hu, Tigang Ning are with Key Laboratory of All Optical Network and Advanced Telecommunication Network, Ministry of Education, Beijing Jiaotong University, Beijing 100044, China (e-mail: 20111015@bjtu.edu.cn; lipei@bjtu.edu.cn; wangjsh@bjtu.edu.cn; zyhu@bjtu.edu.cn; tgning@bjtu.edu.cn;)

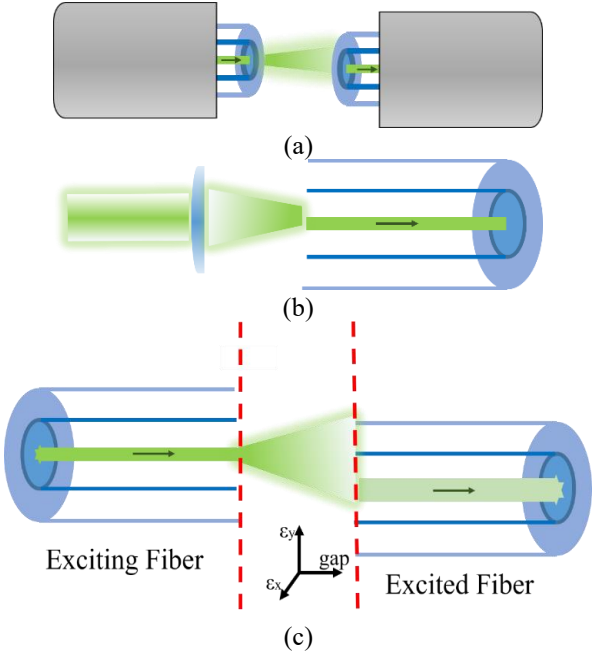

**Fig. 1.** (a) Schematic diagrams of the misaligned free-space coupling structure of two fiber jumper heads, (b) the misaligned free-space coupling structure of a lens and a fiber, (c) and the misaligned free-space coupling structure of uncoated fibers, significantly limiting their practical applications. Furthermore, these techniques can only align the fiber core center and cannot be adjusted based on the resulting mode field. And, the algorithms are not scalable. Specific research on fiber light intensity requires the use of mode decomposition (MD) methods. A combined MD and machine learning (ML)method can be used to simultaneously characterize the light field to observe mode excitation and measure the optical fiber offset distance. Cross-sectional light intensity can be studied, directly measuring the mode field. Further improvements in accuracy can be achieved through increased dataset size and the selection of ML algorithms.

MD is a method to characterize a few-mode light field using the phase and weight of each eigenmode. The number of MD results is only twice the number of eigenmodes. We selected the numerical MD method. The numerical MD method can be implemented using only the light field intensity image obtained by a CCD. Numerical MD methods include deep learning method[20], [21], [22], [23], [24], [25], [26], SPGD method[27], [28], GS method[29], MAA method[30], [31], hybrid GA method[32], MAA-SPGD method[33], and because the MAA-SPGD method has a high decomposition accuracy for the real light field intensity image, it will be used in this paper to study the optical fiber offset excitation situation and measure the distance.

This paper investigates 3D offset displacement in beam alignment systems by integrating MD with ML techniques. We put forward a high-precision 3D measurement approach. This entails devising an extremely straightforward machine learning regression algorithm grounded on MD results. This enables achieving higher alignment accuracy by means of simpler algorithms, more uncomplicated structures, and smaller training sets. Firstly, the mode excitation induced by 3D displacement is systematically analyzed. Building upon this analysis, a hybrid approach MD-ML beam alignment displacement measurement method is developed to measure 3D offset displacement, enabling the monitoring and regulation of micro-deviations in beam alignment links. In simulation studies, the regression model achieves a coefficient of determination ($R^2$) of 0.99 for 3D displacement. In the 3D measurement experiment, the proposed method achieves a coefficient of determination of 0.99 for transverse offsets in the x- and y-directions, and 0.98 for air gap in the z-direction. The RMSE in x-direction, y-direction and z-direction is respectively 0.135 μm, 0.128 μm and 2.42 μm. The time for a single 3D displacement calculation is $4.037 \times 10^{-4}$ seconds. What's more, transverse offset positioning is achieved with a precision of 0.15 μm, though further improvement is limited primarily by motor resolution. Additionally, the mode content excited by spatial light in a few-mode fiber is effectively regulated, yielding an experimental control accuracy with an absolute error of 4.67% in mode power distribution. This research provides a new approach for measuring micro-displacement deviations in environments where the outer cladding wall is difficult to expose and camera ranging is unavailable, significantly improving the feasibility of spatial light alignment measurements in complex and constrained settings.

## II. THEORY

The method proposed in this paper can handle many scenarios in few-mode systems which requiring beam alignment such as adjusting the relative spatial position between two fiber jumper heads as shown in Fig.1 (a), adjusting the relative spatial position between a lens and a fiber as shown in Fig.1 (b), and adjusting the relative spatial position between two optical fibers as shown in Fig.1 (c). Adjusting the relative spatial position of two optical fibers is a common scenario. Therefore, we primarily studied the displacement between the two fibers, as shown in Fig. 1. Here, $\varepsilon_x$ represents the transverse offset between the excited and exciting fibers in the x-direction, $\varepsilon_y$ represents the transverse offset between the two fibers in the y-direction, and gap represents the air gap between the two fibers. This paper investigates the 3D displacement measurement and control based on the change in weight and relative phase of eigenmode caused by the 3D relative offset between the two fibers. Specifically, we use MD to decompose the light field intensity image at a given moment and feed the decomposition result into a trained ML model to obtain the displacement at that moment, which is then measured and controlled. Alternatively, we input the desired mode content and use the trained ML model to obtain the corresponding offset. The desired mode content is then controlled by varying the offset. The training process involves obtaining a series of light field intensity images and their corresponding 3D displacements, decomposing the light field intensity images using MD, and then feeding the decomposition results into the

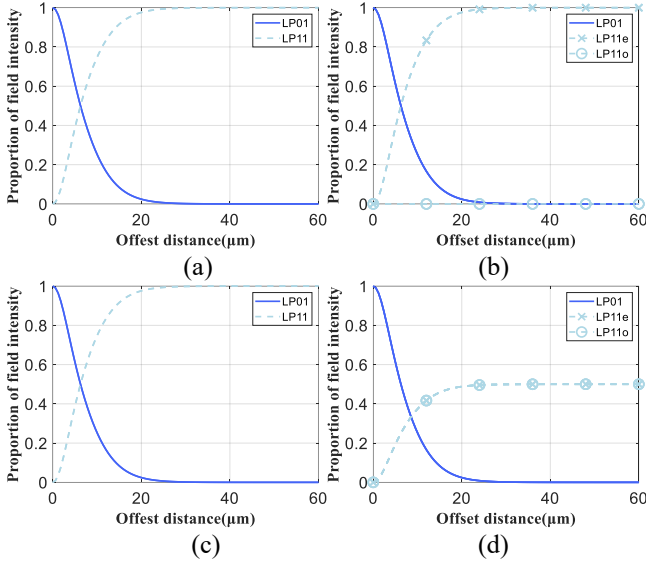

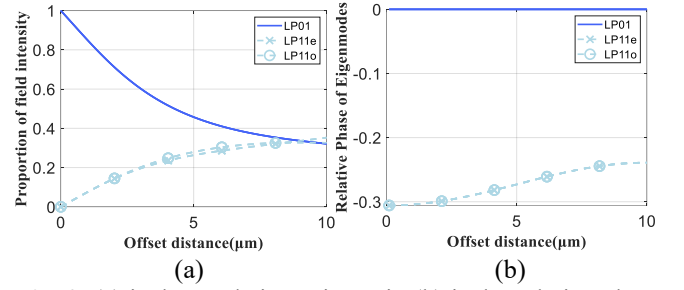

**Fig. 2.** (a) is the mode intensity ratio of the mode group when $\theta=0$, (b) is the mode intensity ratio of the degenerate mode when $\theta=0$, (c) is the mode intensity ratio of the mode group when $\theta=0$, (d) is the mode intensity ratio of the degenerate mode when $\theta=45°$.

**Fig. 3.** (a) is the mode intensity ratio (b) is the relative phase

ML algorithm for training.

Measuring the 3D relative displacement of two optical fibers in space requires the following steps: 1. Light emitted from the exciting fiber propagates in free space; 2. The propagating excitation light is coupled into the excited fiber with a relative offset; 3. After extracting the light field intensity at the output end, the MD algorithm is used to determine the weight and relative phase of each mode (also called mode coefficient); 4. The ML model is used to calculate the 3D relative displacement of the two optical fibers. Below, we will analyze these steps in detail.

*A. Theory of Free-Space Light Propagation*

In free space propagation, since the air gap distance is 10μm -1mm, the Rayleigh-Sommerfeld diffraction method is used for propagation. The field at the *gap* μm distance from the transmitting end is:

$$E_{air}(x_{air}, y_{air}, gap) = \iint E_{src}(x_{src}, y_{src}, 0) h(\Delta x, \Delta y, gap) dx_{src} dy_{src} \quad (1)$$

Among them, $(x_{src}, y_{src}, 0)$ is the source plane, $(x_{air}, y_{air}, gap)$ is the free space plane after propagation, $\Delta x$ is $x_{air} - x_{src}$, $\Delta y$ is $y_{air} - y_{src}$, and the free space transmission kernel function $h(\Delta x, \Delta y, gap)$ is:

$$h(\Delta x, \Delta y, gap) = \frac{e^{ik\sqrt{\Delta x^2 + \Delta y^2 + gap^2}}}{i\lambda\sqrt{\Delta x^2 + \Delta y^2 + gap^2}} \quad (2)$$

Where $k$ is the wave vector, $\lambda$ is the wavelength.

In numerical calculations, the angular spectrum method is generally used to calculate the field at a distance of *gap* position from the transmitting end. The light emitted from the excitation fiber is represented as a superposition of plane waves. Since plane waves only accumulate a phase factor with respect to *z* during propagation, the solution is simple. By superimposing the propagated plane waves, the field distribution at *z* can be obtained. Specifically, a Fourier transform is first performed to obtain the angular spectrum. This is then multiplied by the phase propagation factor at distance *z*. Finally, an inverse Fourier transform is performed to obtain the propagated light field.

*B. Theory of Fiber Offset Coupling*

After propagating through the air, the light field entering the offset excited fiber can be calculated using the mode projection coefficient. Since the modes in the fiber are independent of each other, the projection of each mode of the light field after propagating through the air on the excited fiber can be directly obtained by the overlap integral of the electric field intensity distribution.

$$a_m = \frac{\iint E_{air}(x,y) E_m^*(x,y) dA}{\iint |E_m(x,y)|^2 dA} \quad (3)$$

Where $a_m$ is the mode projection coefficient and $E_m$ is the field distribution of the eigenmode supported by the excited fiber. Therefore, the field input into the fiber can be expressed as a linear superposition of the modes supported by the fiber, as shown in (4).

$$E_{fiber}(x,y) = \sum_m a_m E_m(x,y) \quad (4)$$

Where $E_{fiber}(x, y)$ is the light field distribution input into the optical fiber.

*C. Light Field Mode Information Extraction*

After obtaining a series of light field intensity images, we need to use MD to extract their mode coefficient information for subsequent ML training. Here, we use the MAA-SPGD[33] method for this study. The MAA-SPGD method works as follows: First, use the MAA method to obtain an initial approximate value, then use the SPGD-MD method to obtain a precise value. The MAA algorithm is a mathematical MD method that decomposes the field intensity into an eigenmode matrix and a coefficient matrix. For a fixed optical fiber, the eigenmode matrix is known, and the coefficient matrix contains the light field mode information. To obtain the coefficient matrix, it is only necessary to multiply the inverse of the eigenmode matrix by the field intensity distribution. The SPGD-MD method is an iterative solution. It randomly perturbs the weight phase to create similar gradients in the intensity image. The gradient feedback is used to adjust the mode coefficient to obtain the mode coefficient that maximizes the similarity of the intensity image. Because the matrix analysis method has poor experimental results and the SPGD-MD method is prone to falling into local optimal solutions, the matrix analysis method is used to find an approximate solution, which is used as the initial value for the gradient descent method, thus resolving the problems of both methods.

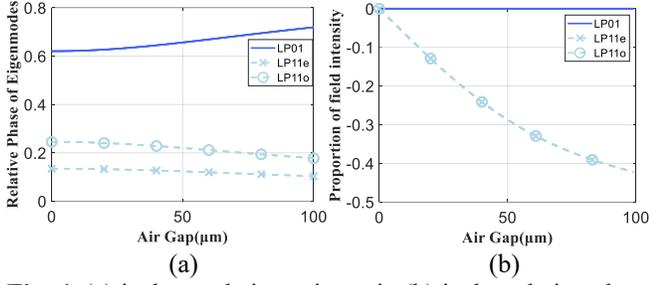

**Fig. 4.** (a) is the mode intensity ratio (b) is the relative phase vs air gap when offset exist.

III. SIMULATIONS AND RESULTS

This section analyzes the possibility of measuring displacements in the *x*-, *y*-, and *z*-direction using mode coefficients through simulation, and simulates the measurement of two-dimensional planar displacements and three-dimensional solid displacements. We used the simplest two-mode fiber for research and simulation calculations. We used a single-mode fiber to interface with a two-mode fiber, with a radius of 6μm and an NA of 0.1076 for this study. Furthermore, since this paper uses linearly polarized modes (LP) for mode decomposition, we will only consider the LP mode from a different perspective below.

*A. Influence of Mode Excitation with Different 3D Displacements on Mode Coefficients*

To ensure rigor, we first need to clarify the influence of displacement in different directions on the eigenmode coefficients excited in the few-mode fiber. This allows us to infer the displacement from the mode coefficients. The displacement includes the offset on the cross-section (*x*-direction, *y*-direction displacement) and the distance of the spatial gap (*z*-direction displacement). Since the offset on the cross-section differs significantly from the longitudinal air gap excitation mechanism, this section discusses the 3D displacement separately in the transverse and longitudinal directions: first, we analyze the influence of the transverse offset on the amplitude and relative phase of each mode, and then analyze the influence of the longitudinal offset on both.

1) *Influence of Transverse Displacement Variation on Eigenmode Coefficients*: When *gap* = 0, $E_{air}$ in (3) is the real amplitude distribution of $LP_{01}$ mode, and $E_{am}$ is also the real amplitude distribution. The calculated mode excitation coefficient is a real number, so the phase is only 0 or π. If the phase of $LP_{01}$ mode in few-mode fiber is taken as the reference phase, the relative phase between modes is only 0 or π. This shows that when the excitation light is a planar Gaussian distribution, the light excited in few-mode fiber has no complex angle. Therefore, when studying the influence of lateral offset on mode coefficients, we mainly focus on the weight changes of each eigenmode. Polar coordinate analysis in the lateral direction can be clearer: the radial distance is r, and the angular angle is $\theta$. We first analyze the case of $\theta$=0, as shown in Fig. 2. In this case, only $LP_{11e}$ mode is excited, and its intensity is related to the x-direction offset. Further consider the lateral offset case of $\theta$=45°, as shown in Fig. 2. Both $LP_{11e}$ and $LP_{11o}$ modes are excited, and their intensity is related to the radial distance *r*. The difference between Fig. 2 (b) and Fig. 2(d) lies in the fact that the change in the offset angle $\theta$ alters the power distribution between the two degenerate modes. Overall, when gap=0 and a lateral offset exists, the mode composition changes with the offset displacement; for different offset angles, the total weight of the same mode group remains unchanged, but the power ratio of its two degenerate modes changes. When the air gap is fixed and not zero, $E_{air}$ is the diverged $LP_{01}$ quasi-spherical wave, while $E_{am}$ remains a real amplitude distribution. We studied the variation of the radial distance *r* of the offset within 0–10 μm under the conditions of gap = 50 μm and $\theta$ = 45°. As shown in Fig. 3, we found that when the gap variation range is small, the effect of offset on the mode weights is similar to that when gap = 0. The effect of offset on the relative phase of the modes is only related to the radial distance *r* and not to the angular angle $\theta$. This is because the quasi-spherical wave, when there is a lateral offset, introduces a laterally correlated phase change at the end face of the excited fiber. Since $LP_{11e}$ and $LP_{11o}$ are excited as a whole, and the $LP_{01}$ and $LP_{11}$ mode groups have circular symmetry, the angular angle $\theta$ has no effect on the relative phase. In summary, when z = 0, the weights of the $LP_{01}$ and $LP_{11}$ mode groups are only affected by the radial distance *r*, $\theta$ determines the ratio between the degenerate modes LP11e and LP11o, and the relative phase of the modes is only 0 or π. When z ≠ 0, the effect of offset on the weights is similar to that when z = 0, while its effect on the relative phase is only related to the radial distance *r*.

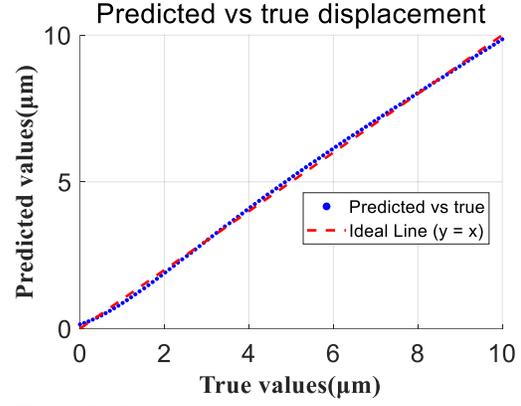

**Fig. 5.** Regression relationship between predicted displacement and actual displacement when only displacement is in the *x*-direction.

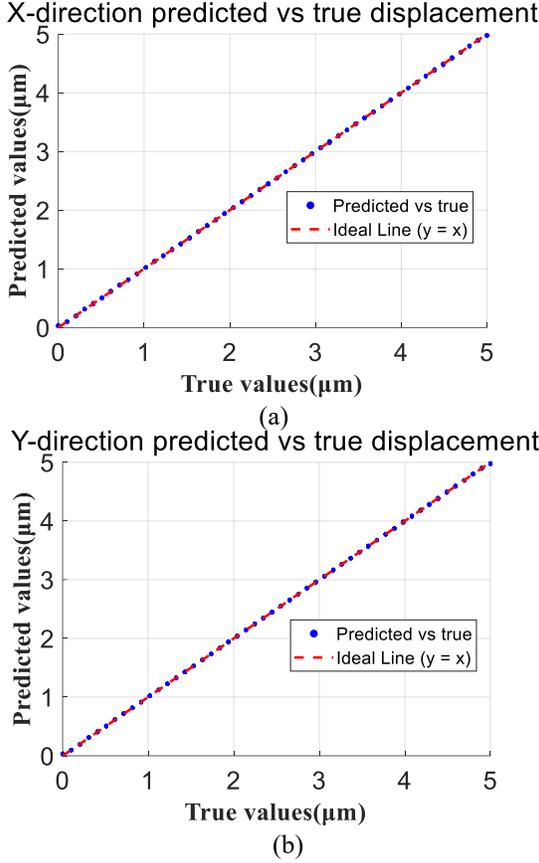

**Fig. 6.** (a) Regression diagram of the predicted displacement in the *x*-direction and the actual displacement; (b) Regression diagram of the predicted displacement in the *y*-direction and the actual displacement when there is displacement in both directions.

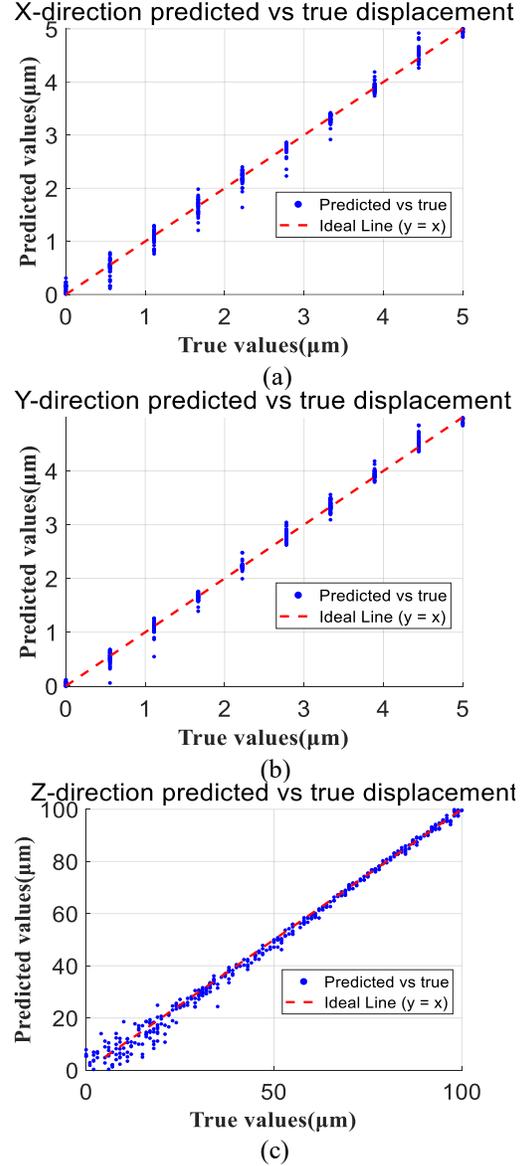

**Fig. 7.** (a) Regression diagram of the predicted displacement in the *x*-direction and the actual displacement; (b) Regression diagram of the predicted displacement in the *y*-direction; (c) Regression diagram of the predicted displacement in the *z*-direction.

2) *Influence of Longitudinal Displacement Variation on Eigenmode Coefficients*: We studied the effect of longitudinal displacement on the eigenmode coefficients in two cases. When offset=0, the air gap has almost no effect on the weights. As shown in Fig. 4, when offset≠0, the air gap causes the excitation light to diverge, which has a certain impact on the projection of the optical field in the few-mode fiber, resulting in a small change in the weights of each eigenmode. Fig. 4 illustrates the effect of the air gap on the relative phase of each eigenmode, showing that the air gap introduces different phase tilts into the diverged excitation light, thus changing the relative phase. It is evident that the air gap primarily affects the relative phase.

From the analysis above, we can see that the information of the transverse offset and the longitudinal air gap is encoded in the weights and relative phase of the eigenmodes. When $z = 0$, the offset can be inferred solely from the mode weights. Therefore, when measuring only the two-dimensional lateral offset, we perform regression based solely on the weights; while when simultaneously measuring the three-dimensional offset, we employ a multi-task learning method (MTL) for regression analysis.

*B. Simulation Measurement of 3D Displacement of Spatial Light*

1) *Simulation Measurement of 1D Spatial Light Displacement*: In the case of offsetting only the *x*-direction, we let the transverse offset displacement vary between 0-10 μm, and simulated and generated a pattern intensity image with its corresponding *x*-direction transverse offset displacement label. Since it performs well in small sample data and has good robustness, the SVR algorithm is used. After performing MD on the pattern intensity image, the content of each mode corresponding to each pattern intensity image is obtained.

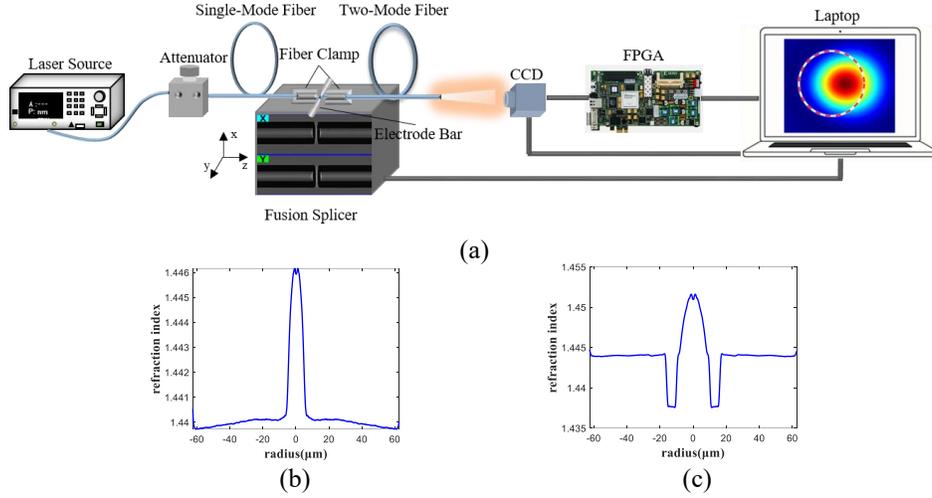

**Fig. 8.** (a) Experimental setup diagram for 3D measurement and regulation of spatial light; (b) The refractive index of the single-mode fiber used in the experiment; (b) The refractive index of the two-mode fiber used in the experiment.

The obtained mode content and displacement label are simultaneously input into the SVR algorithm to obtain a trained SVR model. By performing MD on the test images and substituting them into the SVR model, the *x*-direction transverse offset displacement corresponding to the test image is obtained. The RMSE and $R^2$ are usually used to evaluate the quality of regression. The definitions of RMSE and $R^2$ are shown in (5) and (6) Among them, $o_{true}$ is the observed value and $o_{pred}$ is the predicted value, $n$ is the number of the test data. The relationship between the label and the predicted value is shown in Fig. 5, with an RMSE of 0.1μm and an $R^2$ of 0.9988.

$$RMSE = \sqrt{\frac{1}{n}\sum_{i=1}^{n}(o_{pred} - o_{true})^2} \quad (5)$$

$$R^2 = 1 - \frac{\sum (o_{true} - o_{pred})^2}{\sum (o_{true} - \bar{o}_{true})^2} \quad (6)$$

2) *Simulation Measurement of 2D Spatial Light Displacement*: When offsetting both the *x* and *y* directions, we varied the transverse offset in the *x* and *y* directions between 0 and 5 μm, respectively. The simulation generated a pattern intensity image and its corresponding *x* and *y* transverse offset labels. In 2D measurements, the SVR algorithm is no longer applicable due to the presence of two outputs: the *x* and *y* transverse offsets. We used a random forest algorithm, which is more capable of handling large data sets and complex relationships. After training the model, we performed MD on the test images and fed the results into the random forest model. The relationship between the label and the predicted value is shown in Fig. 6. The RMSE values for the *x*- and *y*-directions are 0.0077μm and 0.0076μm, respectively. The $R^2$ values for the *x*- and *y*-directions are 0.9999 and 0.9999, respectively. Demonstrating the convincing prediction results for the *x* and *y* directions.

3) *Simulation Measurement of 3D Spatial Light Displacement*: We know that if only the 2D offset in the *x* and *y* directions changes, the 2D offset can be calculated using the mode content. When measuring 3D displacement, the phase needs to be included in the calculation. Therefore, we combine the *x* and *y* offsets with the gap for light field calculations, and then use MTL and feature engineering to perform regression. The learning task in MTL consists of two parts: transverse displacement regression and longitudinal displacement regression. RMSE in *x*-direction and RMSE in *y*-direction reach 0.145 μm and 0.117 μm, respectively. $R^2$ in the *x*-direction and $R^2$ in the *y*-direction reach 0.9941 and 0.9953, respectively. In the *z*-direction, because the displacement distance range is relatively large, the grid of the data input to MTL for training is not very fine, resulting in a larger RMSE. RMSE in *z*-direction and $R^2$ in the *z*-direction is 2.692 μm and 0.9930. As shown in Fig. 7, the 3D spatial light offset measurement can also be simulated very well. Furthermore, We classify single-mode optical fibers into two categories based on their Rayleigh length: when z > 50 μm, the RMSE of z is 1.232 μm, and when z < 50 μm, the RMSE of z is 3.400 μm. This demonstrates that the prediction of *gap* becomes more accurate as the distance increases.

## IV. EXPERIMENT AND RESULTS

To achieve 3D spatial light measurement and regulation, an experimental setup as shown in Fig. 8(a) was established. A 1550nm continuous laser beam was emitted from a laser source and entered a single-mode fiber. The laser power in the fiber was then adjusted by an attenuator. A Fujikura FSM-100P fiber splicer was used to achieve 3D relative displacement. Spatial relative displacement is primarily achieved by controlling the position of the fiber clamp using a motor. A CCD camera was placed at the end of the two-mode fiber to capture the mode field transmitted within the fiber. Due to 500 images are required to train the model, we utilized automated control of the fiber splicer and an FPGA-assisted CCD acquisition system to capture the images. Specifically, in the program, the PC controls the 3D displacement of the splicer and, upon completion, sends commands to the FPGA. Upon receiving the commands from the PC, the FPGA invokes a program on the chip to control the CCD to automatically capture images. Finally, the CCD saves the images to the PC. The program then continues to execute the next splicer action. During each splicer operation, we used a camera capable of imaging the fiber cladding sidewalls using a dual-view imaging method to obtain the 3D relative displacement. In our experiments, we used a single-mode fiber and a two-mode fiber. The refractive index of the two-mode fiber used in the experiment is shown in Fig. 8(b). The modes in the two-mode fiber are $LP_{01}$, $LP_{11e}$, and $LP_{11o}$, respectively. In our experimental study, we obtained 100 sets of light intensity images with displacement only in the *y*-direction to investigate 1D displacement. We also obtained 676 sets of light intensity images with displacements in both *x* and *y* directions to investigate 2D displacement. Furthermore, we obtained 3380 sets of light intensity images with displacements in 3D displacement to investigate. We also obtained 676 sets of light intensity images with displacements in both *x* and *y* directions to investigate 2D displacement. Furthermore, we obtained 3380 sets of light intensity images with displacements in 3D displacement to investigate. Finally, using these measurement results, we regulated the relative displacement of spatial light.

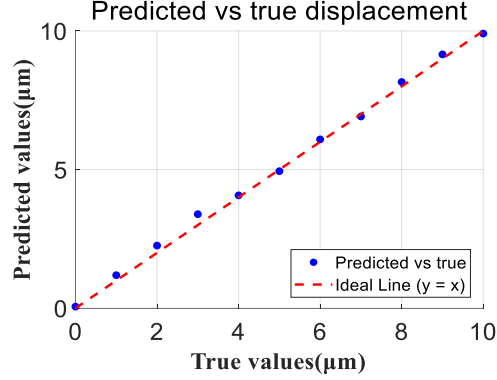

**Fig. 9.** Regression of the predicted and the actual displacement with displacement only in the *x*-direction.

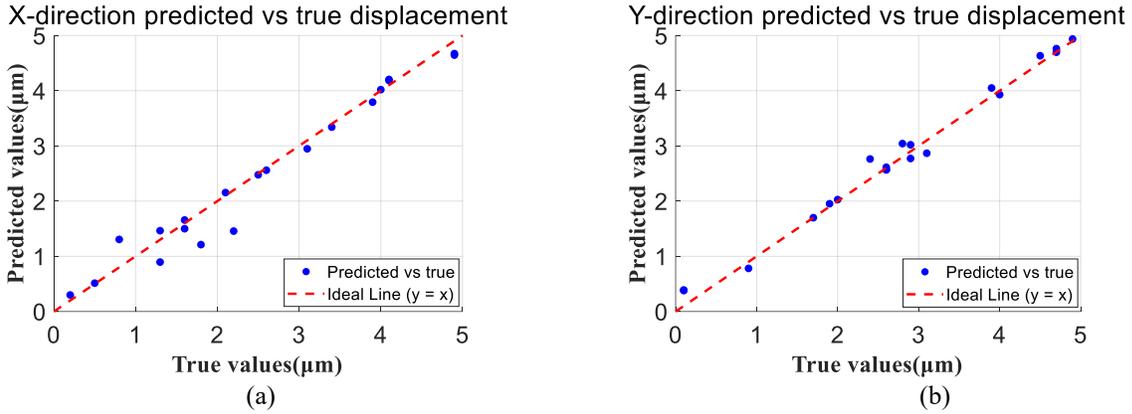

**Fig. 10.** (a) Regression diagram of the predicted displacement in the *x*-direction and the actual displacement; (b) Regression diagram of the predicted displacement in the *y*-direction and the actual displacement when there is displacement in both directions in the experiment.

*A. Measurement and Regulation of Spatial Light Displacement*

1) *Experimental Measurement of 1D Spatial Light Displacement*: This section presents experimental results for two optical fibers with only *y*-direction displacement. In this *y*-direction-only displacement experiment, we controlled the air gap to 10 μm and captured 100 light field intensity images with air gaps ranging from 0 to 10 μm in the *x*-direction. After centering and denoising these images, the MD algorithm was used to obtain the relative intensity power of each mode. These power values were then substituted into the SVR algorithm for training, yielding a trained SVR model. The intensity distribution at 11 test points was also subjected to MD, and the results were substituted into the SVR model to determine the *y*-direction displacement corresponding to these 11 test points. The displacement calculated using this method is shown in Fig. 9. As shown in Fig. 9, the displacement measurements obtained using only the 100 MD results as a training set and basic SVR testing are generally consistent with the true displacement values. The RMSE of these 11 displacement measurements is calculated to be 0.1766 μm. The $R^2$ of these measurements is 0.9969.

2) *Experimental Measurement of 2D Spatial Light Displacement*: This section presents the experimental results of 2D offset measurement of two optical fibers in the *x* and *y* directions. In this 2D offset measurement experiment, we controlled the air gap to 10 μm and captured 26 light field intensity images in each direction, totaling 676 images. Image processing is similar to the 1D offset measurement, differing in that a random forest algorithm is used to train the MD results in 2D. After obtaining the trained random forest model, the MD results of the light intensity distribution at 20 test points were substituted into the random forest model to determine the relative offset positions of these 20 points on the cross-section. Calculations show that the RMSE of the offset between these 20 sets of offset distance measurements and the images captured by the splicer camera is 0.2782 μm in the *x*-direction and 0.1593 μm in the *y*-direction. And the $R^2$ is 0.9607 in the *x*-direction and 0.9868 in the *y*-direction.

3) *Experiment Measurement of 3D Spatial Light Displacement*: This section presents the experimental results of 3D displacement measurement using two optical fibers. In this 3D displacement measurement experiment, we selected 5 points in the z-direction and collected 26 points each in the x and y directions, totaling 3380 images. With variations in the *z*-direction, we needed to incorporate the phase of the eigenmodes into the parameter range and trained the model using the MTL method. After obtaining the trained MTL model, we used 200 points for prediction. The prediction results show that the RMSE between these 200 sets of displacement distance measurements and the true values is 0.135 μm in the *x*-direction and 0.128 μm in the *y*-direction. The RMSE in *z*-direction is 2.42 μm. The $R^2$ in *x*-direction, *y*-direction and *z*-direction is 0.9924, 0.9920 and 0.9827, respectively. The time for a single 3D displacement calculation is $4.037 \times 10^{-4}$ seconds.

*B. Regulation of Spatial Light Offset and Mode Distribution*

In our experiment, we not only measured the spatial light offset but also controlled it with a gap of 30 μm. After the computer calculated the offset, the offset position was adjusted in a single step based on the results to achieve the desired offset. To adjust the offset to any arbitrary position, we selected (4.9, 2.0) for adjustment and located it at (4.8, 1.8), with an error of approximately 0.15 μm. We also implemented an offset adjustment to locate any arbitrary mode content. With a gap of 30 μm, we selected positions where the $LP_{01}$, $LP_{11e}$, and $LP_{11o}$ contents were 72%, 25%, and 3%, respectively, and performed a regression search. The result was located at (1.59, 2.29). A table lookup revealed that the intensity image contents of each mode at this point were 68%, 22%, and 10%, respectively. We calculated the absolute error using the (7). The absolute error was 4.67%.

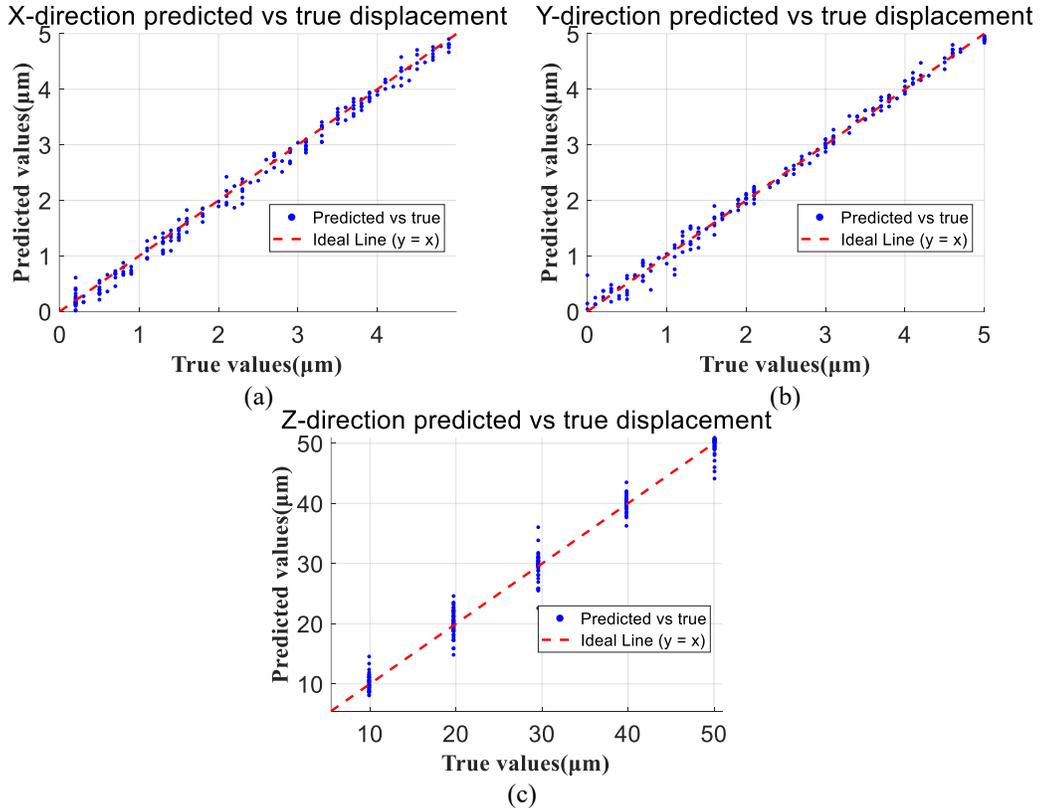

**Fig. 11.** (a) Regression diagram of the predicted displacement in the *x*-direction and the actual displacement; (b) Regression diagram of the predicted displacement in the *y*-direction; (c) Regression diagram of the predicted displacement in the *z*-direction.

$$MAE = \frac{1}{n}\sum_{i}\left| p_i^{\text{true}} - p_i^{\text{pred}} \right| \quad ()$$

Where $p_i^{\text{true}}$ is the actual power distribution percentage and $p_i^{\text{pred}}$ is the predicted power percentage.

## V. Discussion and Results

This paper investigates a method for measuring 3D displacement of optical fibers using single-image sensing of optical fiber cross sections based on MD. This method eliminates the need for measurement of the fiber's outer wall, making it unaffected by the camera's field of view. Single-image sensing of the cross-section enables unimpeded measurement in spatially constrained environments and can also be used for sensing in communications. This approach overcomes the limitations of traditional optical fiber 3D displacement measurement, adding a vast array of previously unavailable applications with a simpler structure. Furthermore, the ease of computer processing of MD and the ease of processing the very small number of values resulting from MD in ML significantly reduce the computational cost of this 3D displacement measurement solution compared to other approaches.

Our experimental results are less than surprising, primarily due to two factors. First, our focus is on proposing a new solution for 3D fiber measurement. Therefore, we did not tune or optimize the ML method, and we simply selected an appropriate ML algorithm. Second, after the excitation light is scattered in space and projected into a few-mode fiber at a certain offset, due to the fiber's structural characteristics, only the modes projected into the core can continue to propagate. Many modes projected into the cladding are rapidly lost, and the loss of these cladding modes results in a loss of light field information. However, our measurements are closely tied to light field information, which reduces measurement accuracy. Therefore, the more modes a fiber supports and the greater the proportion of transmissive modes projected into the core, the higher the measurement accuracy. Our current measurements were demonstrated using a single two-mode fiber, representing the simplest and least accurate case. Therefore, although the current accuracy is limited to submicron levels, the improved accuracy of this approach is very promising. Our current results show that even without adjusting ML parameters and using only two-mode fiber calculations, we can achieve an average accuracy of 0.0076 μm in the $x$ and $y$ directions in simulations on the $x$-$y$ plane. Furthermore, experiments have proven this method is very simple, with results available within $4.037 \times 10^{-4}$ seconds. However, this approach also has its limitations. In environments with significant external interference, larger datasets may be required to model, distinguish, and denoise different types of interference.

This paper combines MD with ML to conduct a detailed study of the 3D spatial displacement in a few-mode spatial optical system. The study encompasses an analysis of mode excitation under different 3D displacement conditions, precise measurement of 3D displacements, and active regulation of both spatial offset displacement and few-mode light field content. The method innovatively utilizes methods that investigate the intrinsic properties of the optical field within the fiber to achieve the localization of the 3D displacement. For 3D displacement measurement, we implemented random forest for 2D measurement, and an MTL framework for full 3D reconstruction, achieving R² values exceeding 0.99 in all cases. Experimental validation confirmed the robustness of the approach, yielding R² > 0.99 for 2D and R² > 0.98 for 3D measurements. Furthermore, we achieved high-precision spatial offset regulation with a positional deviation of only 0.15 μm from the target. The methodology was extended to control the few-mode field content, resulting in a low absolute error of 4.67%. By shifting the measurement paradigm from reliance on cladding surface features to direct analysis of optical field intensity distributions, this work establishes a novel framework for 3D spatial light characterization. This study proposes a new method for the 3D measurement of spatial light from the perspective of optical fiber eigenmode coefficients calculation, which changes the measurement of spatial light offset position from detecting the outer wall of the cladding to detecting the light field intensity, expanding the application of fiber alignment in fields such as spatial light alignment, communication perception, and three-dimensional distance sensors.